\documentclass{article}

\usepackage{arxiv}

\usepackage[utf8]{inputenc} 
\usepackage[T1]{fontenc}    
\usepackage{url}            
\usepackage{booktabs}       
\usepackage{amsfonts}       
\usepackage{amsmath,bm,upgreek}
\usepackage{nicefrac}       
\usepackage{microtype}      
\usepackage{graphicx}
\usepackage[square,numbers]{natbib}
\usepackage{hyperref}       
\usepackage{cleveref}       
\usepackage{doi}
\usepackage{subfig}

\DeclareMathOperator{\T}{\mathsf{T}}

\newcommand{\pyb}{p(\mathbf{y|b})}
\newcommand{\pybb}{p(\mathbf{y|b}+\Delta\mathbf{b})}

\title{Decision-oriented two-parameter Fisher information sensitivity using symplectic decomposition}

\date{October 12, 2022}

\author{ \href{https://orcid.org/0000-0001-8323-7406}{\includegraphics[scale=0.01]{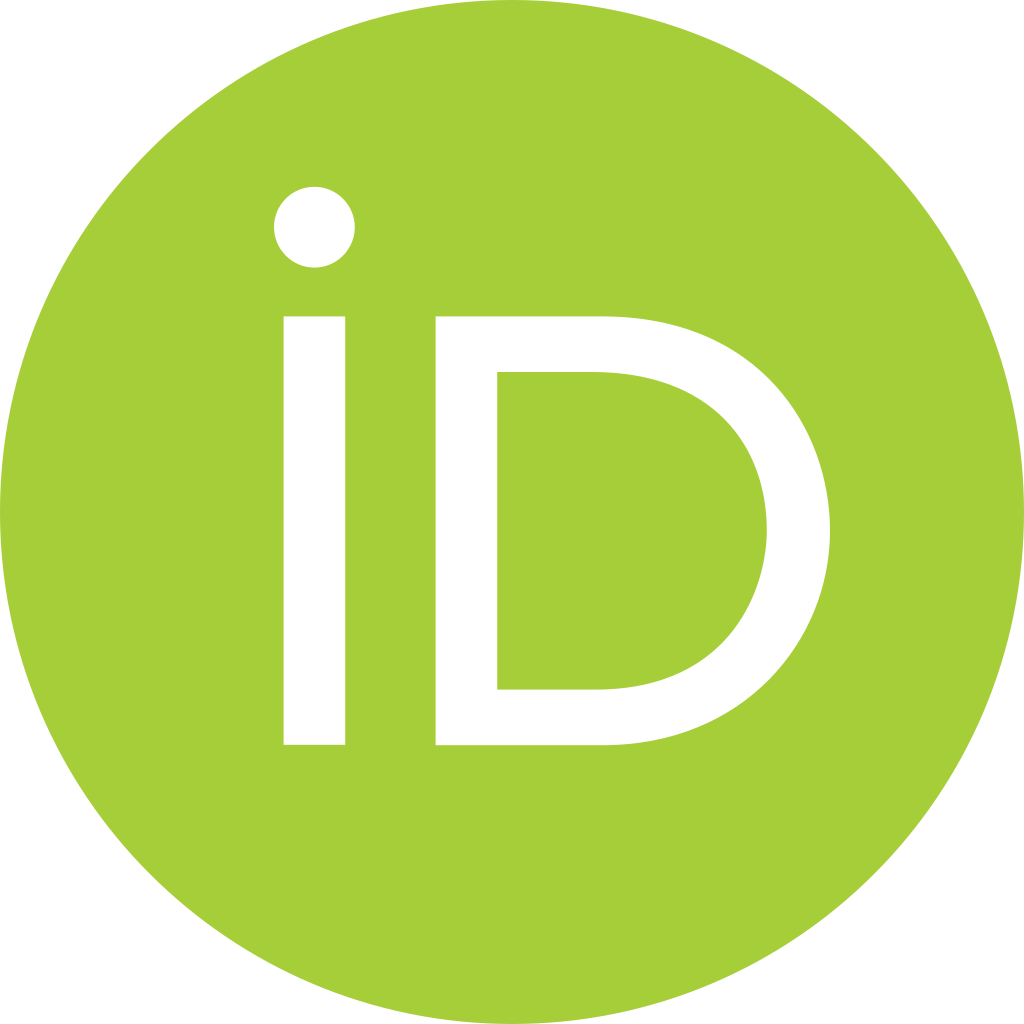}\hspace{1mm}Jiannan Yang}
    \\
  	Department of Engineering\\
	University of Cambridge\\
	Trumpington Street, Cambridge CB2 1PZ, UK\\
	\href{mailto:jy419@cam.ac.uk}{jy419@cam.ac.uk}\\
}

\renewcommand{\shorttitle}{\textit{Symplectic Fisher Sensitivity}}

\hypersetup{
pdftitle={Decision-oriented two-parameter Fisher information sensitivity using symplectic decomposition},
pdfsubject={},
pdfauthor={Jiannan Yang},
pdfkeywords={decision under uncertainty, probabilistic sensitivity, Williamson’s theorem, symplectic eigenvalue, conjugate parameters},
}

\begin{document}
\maketitle

\begin{abstract}
The eigenvalues and eigenvectors of the Fisher information matrix (FIM) can reveal the most and least sensitive directions of a system and it has wide application across science and engineering. We present a symplectic variant of the eigenvalue decomposition for the FIM and extract the sensitivity information with respect to two-parameter conjugate pairs. The symplectic approach decomposes the FIM onto an even dimensional symplectic basis. This symplectic structure can reveal additional sensitivity information between two-parameter pairs, otherwise concealed in the orthogonal basis from the standard eigenvalue decomposition. The proposed sensitivity approach can be applied to naturally paired two-parameter distribution parameters, or a decision-oriented pairing via re-grouping or re-parameterization of the FIM. It can be utilised in tandem with the standard eigenvalue decomposition and offer additional insight into the sensitivity analysis at negligible extra cost.
\end{abstract}

\keywords{decision under uncertainty \and probabilistic sensitivity \and Williamson’s theorem \and symplectic eigenvalue \and conjugate parameters}

\section{Introduction}
\label{sec:1}
\subsection{Background}
\label{sec:1.1}
Sensitivity analysis is an integral part of mathematical modelling, and in particular a crucial element of decision making in presence of uncertainties. The Fisher information, first introduced by Fisher \cite{Fisher1922On} and is widely used for parameter estimation and statistical inference, has found increasing application in many area of science and engineering for probabilistic sensitivity analysis. For example, the Fisher Information Matrix (FIM) has been applied to the parametric sensitivity study of stochastic biological systems \cite{Gunawan2005Sensitivity}; the FIM is used to study sensitivity, robustness, and parameter identifiability in stochastic chemical kinetics models \cite{Komorowski2011Sensitivity}; through link with relative entropy, the Fisher information is used to assess the most sensitive directions for climate change given a model for the present climate \cite{Majda2010Quantifying}; used in conjunction with the principle of Max Entropy, the FIM is used to identify the pivotal voters that could perturb the collective voting outcomes in social systems \cite{Lee2020Sensitivity}; and more recently in \cite{Yang2022Digital} the Fisher information have been proposed as one of the process-tailored sensitivity metrics for engineering design. Despite the wide scope, the applications mentioned above all utilise the spectral analysis of the FIM, i.e., the eigenvalues and eigenvectors of the FIM reveal the most and least sensitive directions of the system.

In this paper, we apply a symplectic spectral analysis of the FIM, and demonstrate that the resulted symplectic eigenvalues and eigenvectors are oriented towards a better decision support by extracting sensitivity information with respect to two-parameter pairs (e.g., the mean and standard deviation of a Normal distribution). Consider a general function $\bf y=h(x)$, the probabilistic sensitivity analysis characterise the uncertainties of the output $\bf y$ that is induced by the random input $\bf x$ \cite{oakley_probabilistic_2004, oakley_decision-theoretic_2009}. When the input can be described by parametric probability distributions, i.e., $x \sim p(\mathbf{x|b})$, the FIM can then be estimated as the covariance matrix of the random gradient vector ${\partial\ln\pyb}/{\partial \mathbf b}$, with the $jk^{\text{th}}$ entry of the Fisher Information Matrix (FIM) as (e.g., \cite{Yang2022Digital}):
\begin{equation} \label{eq:1}
	F_{jk}=\int \frac {\partial p(\mathbf{y|b})}{\partial b_j} \frac {\partial p(\mathbf{y|b})}{\partial b_k} \frac {1}{p} d\mathbf y = \mathbb E_Y\left[ \frac { \partial \ln p}{\partial b_j} \frac {\partial \ln p}{\partial b_k} \right]
\end{equation}
The eigenvalues of the FIM represent the magnitudes of the sensitivities with respect to simultaneous variations of the parameters $\mathbf{b}$, and the relative magnitudes and directions of the variations are given by the corresponding eigenvectors. 

The FIM depends on the parametrization used. Suppose $b_j=g_j(\theta_i), \: i=1,2,\dots,s$, then the FIM with respect to the parameter $\bm{\uptheta}$ is \cite{Lehmann1998Theory}:
\begin{equation} \label{eq:2}
    \mathbf{F}(\bm{\uptheta})=\mathbb{J}^\mathsf{T}\mathbf{F}(\mathbf{b})\mathbb{J} 
\end{equation}
where $\mathbb{J}$ is the Jacobian matrix with $\mathbb{J}_{ji}={\partial b_j}/{\partial \theta_i}$.

It should be noted that the sensitivity analysis based on FIM is fundamentally different from the commonly used variance based analysis \cite{saltelli_global_2008}. The Fisher sensitivity examines the perturbation of the entire joint probability density function (PDF) of the output, more specifically, the entropy of output uncertainty. Moreover, the sensitivity measure from the Fisher analysis are the eigenvectors, which can be regarded as the principal directions for a simultaneous variation of the input parameters. This is in contrast to variance based ranking where it is assumed that the uncertainty of the input factors can be completely reduced to zero \cite{oakley_probabilistic_2004}. As pointed out in \cite{Yang_SAframework_2022}, using principal sensitivity directions is based on a pragmatic view that given a finite budget to change the parameters, maximizing the impact on the output follows the principal sensitivity directions, which tend to be a simultaneous variation of the parameters because their effect on the output are likely to be correlated. The constrained maximization view also leads to the symplectic eigenvectors in a symplectic basis as discussed in Section \ref{sec:3.1}. 

The Fisher sensitivity is based on partial derivatives, but it is different from the derivative based global sensitivity measure \cite{Sobol_derivative_2009} which is defined as the integral of the squared derivatives of the function output. The Fisher information, on the other hand, is defined as the variance of the partial derivatives of the log probability of the uncertain function output, as seen in Eq \ref{eq:1}. And this differentiation is with respect to the distribution parameters of the uncertain input, not with respect to the uncertain variables themselves. Therefore, the Fisher sensitivity examines the impact of the perturbation of the input probability distribution, and as the input distributions are often estimated from data, it is equivalent to assess which uncertain dataset to be focused on.  Sensitivity index based on modification of the input PDF has been proposed in \cite{Lemaitre_densitySA_2015} for reliability sensitivity analysis, where the input perturbation is derived from minimising the probability divergence under constraints. In contrast, we consider parametric uncertain inputs in this paper to form the Fisher information matrix (FIM) and the resulted eigenvectors provide the principal directions for the input perturbation.     

Many widely applied parametric distributions are in the two-parameter families, e.g., the location-scale families including the Normal distribution and Gamma distribution. Although the Fisher sensitivity is with respect to these distribution parameters $\bf b$, the quantities of interest for decision-making are ultimately the uncertain variables $\mathbf{x}$ themselves, e.g., to rank the relative importance of $\bf x$. We will demonstrate in this paper that the symplectic decomposition of the FIM identifies the influential two-parameter pairs, or equivalently the corresponding variables, and can be used in tandem with the standard eigenvalue decomposition for a better decision support. 
\subsection{A motivating example}
\label{sec:1.2}
As a motivating example, we consider an engineering design problem under uncertainties. Consider a simple cantilever beam where the Young’s modulus $E$ and the length $L$ are uncertain, i.e., $\mathbf{x}=(E,L)$, and the uncertainties can be described by Normal distributions with $E \sim \mathcal{N} (\mu_1=69e9,\: \sigma_1^2=11.5e9^2 )$ and $L \sim \mathcal{N}(\mu_2=0.45, \: \sigma_2^2 = 0.045^2)$. To keep it analytically tractable for this motivating example, we assume a trivial function $\mathbf{y=x}$ (a random vibration problem considered in Section \ref{sec:4}). Assuming the two random variables are independent, the Fisher information matrix in this case is a diagonal matrix \cite{Cover2006Elements}: 
\begin{equation} \label{eq:3}
      \mathbf{F}(\mu_1,\sigma_1,\mu_2,\sigma_2)=\text{diag}\left(\sigma_1^{-2},2\sigma_1^{-2},\sigma_2^{-2},2\sigma_2^{-2}\right) 
\end{equation}

The eigenvalues, the diagonal entries of the FIM in Eq \ref{eq:3} in this case, and the corresponding eigenvectors then provide the sensitivity information of the uncertain output $\bf y$ with respect to (w.r.t) the input parameter vector $\mathbf{b}=(\mu_1,\sigma_1,\mu_2,\sigma_2 )$. More specifically, the sensitivity of the entropy of the random output as to be discussed in Section \ref{sec:3}. 

For practical utilisation of the Fisher sensitivity information, there are two issues that need to be addressed and that motivates our research in this paper. First, the FIM needs to be normalised. On one hand, the un-normalised FIM given in Eq \ref{eq:3} tends to be ill conditioned. For example, the conditional number is in the order of  $10^{22}$ given that $\sigma_1=11.5e9$ and $\sigma_2=0.045$. On the other hand, as the Young’s modulus $E$ and the length $L$ are of different units, the FIM needs to be normalised so that the sensitivities w.r.t the different parameters are comparable. One option is to consider sensitivity w.r.t a percentage change of the parameter and this is called proportional \cite{Yang2022Digital} or logarithmic \cite{Pantazis2013Parametric}       normalised FIM. Normalization is equivalent to a re-parametrization. In the case of proportional normalization, the change of parameter is $b_j=\bar{b}_j\theta_j$ with $\bar{b}_j$ the nominal value for normalization, and the Jacobian matrix in Eq \ref{eq:2} is just a diagonal matrix with $\bar{b}_j$ on the diagonal. 

However, the proportional normalisation might provide unrealistic sensitivity information for practical applications. For example, unless the probability distribution of the input variables is far from the real distribution, it is most likely that the change of the mean should be within one or two standard deviations. The FIM can instead be normalised by the standard deviations, which implies that the allowable range of the mean value is limited to a local region and it is quantified by the standard deviation. Normalising the FIM from Eq \ref{eq:3} by the standard deviations, i.e., $\bar{b}_j$ equal to the corresponding $\sigma$, we have the normalised FIM as:
\begin{equation} \label{eq:4}
    \mathbf{F}_{\text{nor}}=\text{diag}(\sigma_1^{2}\sigma_1^{-2},2\sigma_1^{2}\sigma_1^{-2},\sigma_2^{2}\sigma_2^{-2},2\sigma_1^{2}\sigma_2^{-2})= \text{diag}(1,2,1,2)
\end{equation}

where it is evident the condition number of the normalised FIM is much smaller. However, the $\mathbf{F}_\text{nor}$ in Eq \ref{eq:4} has repeated eigenvalues, and as a result, the corresponding eigenvectors are not unique. Although the situation with repeated eigenvalues might seem extreme, as to be seen with more examples, the eigenvalues of the normalised FIM tend to be of similar magnitudes. In other words, the sensitivity information has been compressed by normalization (in exchange for better conditioning). As we shall see, the symplectic decomposition of the FIM has a unique symplectic structure, and that tends to mitigate this issue by making the sensitivity information for different variables more distinctive (by pairing the parameters). 

The second issue with the Fisher sensitivity with respect to the distribution parameters is the gap to decision making. The purpose of the sensitivity analysis is to identify the influential variables so that informed decisions can be made. Although it is possible to make changes to the mean and standard deviation independently, the quantities of interest are ultimately the variables themselves, i.e. $E$ and $L$ in this case. As to be demonstrated, the symplectic approach would naturally put parameters in pairs, e.g., $(\mu, \: \sigma)$ as a conjugate pair for random input with Normal distribution, and provide more direct support for decision making. It is noted in passing that even the true distribution of the uncertain input is not Normal, a common practice is still to use mean and standard deviation as the summary statistic for the dataset at hand. As a result, the two-parameter pair sensitivity proposed in this paper still applies. 
\subsection{Summary and paper outline}
\label{sec:1.3}
In summary, the use of the Fisher information as a sensitivity measure has wide application across science and engineering. Nevertheless, practical issues can hinder the translation of the sensitivity information into actionable decisions. In this paper, we propose a new approach using the symplectic decomposition to extract the Fisher sensitivity information. The symplectic decomposition utilises Williamson’s theorem \cite{Nicacio2021Williamson, Williamson1936On} which is a key theorem in Gaussian quantum information theory \cite{Pereira2021Symplectic}. Originated from Hamiltonian mechanics, the symplectic transformations preserve Hamilton’s equations in phase space \cite{Arnol'd1989Mathematical}. In analogy to the conjugate coordinates for the phase space, i.e. position and momentum, we regard the input parameters as conjugate pairs and use symplectic matrix for the decomposition of the Fisher information matrix (FIM). The resulted symplectic eigenvalues of large magnitude, and the corresponding symplectic eigenvectors of the FIM, then reveal the most sensitive two-parameter pairs. 

It should be noted that the proposed symplectic decomposition is only applicable for parameter space of even dimensions, i.e., $\mathbf{b} \in \mathbb{R}^{2n}$, and the requirement that the parameters can be regarded as two-parameter pairs. For two-parameter family of probability distributions, such as the widely used location-scale families, there is a natural pairing of the parameters. For other cases, a decision-oriented pairing might be needed. For example, a re-parametrization with respect to (w.r.t) the mean and standard deviation, or two moments of the random variables, using Eq \ref{eq:2} would transform the Fisher information matrix (FIM) into even dimensions. Once the FIM is obtained w.r.t parameters of even dimensions, it is envisaged that the proposed symplectic decomposition is best utilised in tandem with the standard eigenvalue decomposition for sensitivity analysis using the FIM. This offers additional insight into the sensitivity analysis, and as the main computational burden is often at estimating the FIM, at negligible extra cost. 

In what follows, we will first review the approach of symplectic decomposition using the Williamson’s theorem in Section \ref{sec:2}. The details of finding the symplectic eigenvalues and eigenvectors are given in the \emph{Supplementary Material} together with the corresponding Matlab script. In Section \ref{sec:3}, we give a theoretical comparison between the symplectic decomposition and the standard eigenvalue decomposition, in terms of the sensitivity of entropy and also from optimization point of view using trace maximization. A benchmark study is conducted in Section \ref{sec:bench}, where the similarity and difference between the Fisher sensitivity and the main effects indices used in variance based analysis are discussed. In Section \ref{sec:4}, a numerical example using a simple cantilever beam is used to demonstrate the effect of symplectic decomposition. Concluding remarks are given in Section \ref{sec:5}. 
\section{Symplectic decomposition}
\label{sec:2}
From elementary linear algebra, we know that a real symmetric matrix $\mathbf{F}$ can be diagonalized by orthogonal matrices:
\begin{equation} \label{eq:5}
    \bf Q^{-1}FQ = \Lambda 
\end{equation}

where $\mathbf{Q}$ is the orthogonal eigenvector matrix, i.e. $\mathbf{Q}^{\T}=\mathbf{Q}^{-1}$, and $\mathbf{\Lambda} = \text{diag}(\lambda_1,\lambda_2,\dots)$ contains the real eigenvalues. And the solution to Eq \ref{eq:5} can be solved using the standard eigenvalue equation:
\begin{equation} \label{eq:6}
    \bf{FQ = Q\Lambda}, \quad  \text{with} \quad  \det(\mathbf{F}-\lambda\mathbf{I})=0 
\end{equation}

The Williamson’s theorem provides us with a symplectic variant of the results above. Let $\mathbf{F \in \mathbb{R}^{2n \times 2n}}$ be a symmetric and positive definite matrix, the Williamson’s theorem says that $\mathbf{F}$ can be diagonalized using symplectic matrices \cite{Nicacio2021Williamson, Gosson2006Symplectic}: 
\begin{equation} \label{eq:7}
    \bf S^{\mathsf{T}}FS = \hat{D} =
    \begin{bmatrix}
         \mathbf{D}&  \\
   		 & \mathbf{D} 
    \end{bmatrix}	
\end{equation}

where $\mathbf{D} = \text{diag}(d_1,d_2,\dots,d_n)$ is a diagonal matrix with positive entries ($d_j$ maybe zero if $\mathbf{F}$ is semidefinite). The $d_j, \: j=1,2,\dots,n$ are said to be the symplectic eigenvalues of matrix $\mathbf{F}$ \cite{Bhatia2015On} and are in general not equal to the eigenvalues given in Eq \ref{eq:5}. The matrix $\mathbf S = [\mathbf{u}_1,\dots,\mathbf{u}_n,\mathbf{v}_1,\dots,\mathbf{v}_n]$ is a real symplectic matrix that satisfies the condition: 
\begin{equation} \label{eq:8}
    \bf S^{\mathsf{T}}JS = J, \quad  \text{where} \quad  \mathbf{J}= 
    \begin{bmatrix}
         \mathbf{0_n} & \mathbf{I_n}  \\
   		 -\mathbf{I_n} & \mathbf{0_n}
    \end{bmatrix}	
\end{equation}

The matrix $\mathbf{I}_n$ is the identity matrix of size $n$ and the matrix $\mathbf{J}$ is itself a symplectic matrix. 
From Eq \ref{eq:8}, we have the following expression for $\mathbf{S}^{-\T}$ , i.e. the inverse transpose: 
\begin{equation} \label{eq:9}
    \bf S^{\mathsf{T}}JS = J \rightarrow 
    \bf JS=S^{-\mathsf{T}}J \rightarrow
    \bf JSJ^{-1}=S^{-\mathsf{T}}
\end{equation}

Substitute the expression for $\mathbf{S}^{-\T}$ into Eq \ref{eq:7}, we have:
\begin{equation} \label{eq:10}
    \bf FS=JSJ^{-1}
    \begin{bmatrix}
         \mathbf{D}&  \\
   		 & \mathbf{D} 
    \end{bmatrix}	
   = JS
    \begin{bmatrix}
          & -\mathbf{D} \\
   	\mathbf{D} & 
    \end{bmatrix}
\end{equation}
where the identity $\mathbf{J}^{-1}=-\mathbf{J}$ is used. Taking the analogy with Eq \ref{eq:6}, the matrix $\mathbf{S}$ is called the symplectic eigenvector matrix of $\mathbf{F}$.
From Eq \ref{eq:10}, we can see that each symplectic eigenvalue $d_j$ corresponds to a pair of eigenvectors $\mathbf{u}_j, \: \mathbf{v}_j \in \mathbb{R}^{2n}$:
\begin{equation} \label{eq:11}
    \mathbf{Fu}_j = d_j\mathbf{Jv}_j; \quad 
    \mathbf{Fv}_j = -d_j\mathbf{Ju}_j
\end{equation}

These eigenvector pairs can be normalized so that they form an orthonormal basis for the symplectic vector space:
\begin{equation} \label{eq:12}
    \mathbf{u}_i^{\mathsf{T}} \mathbf{J} \mathbf{v}_j= \delta_{ij}, \quad 
    \text{for} i,j=1,2,\dots,n
\end{equation}
From Eq \ref{eq:10}, the symplectic decomposition of $\mathbf{F}$ can also be written as:
\begin{equation} \label{eq:10b}
    \bf F=JSJ^{-1}
    \begin{bmatrix}
         \mathbf{D}&  \\
   		 & \mathbf{D} 
    \end{bmatrix}
    	S^{-1}
   = JSJ^{-1}
    \begin{bmatrix}
         \mathbf{D}&  \\
   		 & \mathbf{D} 
    \end{bmatrix}
    	\left(JSJ^{-1}\right)^{\T}
    =M
    \begin{bmatrix}
         \mathbf{D}&  \\
   		 & \mathbf{D} 
    \end{bmatrix}
    	M^{\T}
\end{equation}
where the expression in Eq \ref{eq:9} is substituted for $\bf S^{-1}$ in the second to last step. From Eq \ref{eq:10b}, it is clear that the symmetric form of the matrix $\mathbf F$ is preserved. In addition, as the determinant of symplectic matrices are always one, from Eq \ref{eq:7}, the determinant of the FIM can be expressed as the product of its squared symplectic eigenvalues: 
\begin{equation} \label{eq:10c}
   \det( \mathbf {S^{\mathsf{T}}FS}) = \det( \mathbf S^{\mathsf{T}})\det(\mathbf F)\det(\mathbf S) =  \det(\mathbf F) =  \Pi_{j=1}^n{d_j^2}
\end{equation}
As the determinant of the FIM is also equal to the product of its standard eigenvalues, i.e., $\det \mathbf{F} = \Pi_{j=1}^{2n}\lambda_j$, Eq \ref{eq:10c} implies the total sensitivity volume is conserved in the symplectic decomposition. The volume is in terms of the relative entropy and more details will be given in Section \ref{sec:3.2}. \\
It should be noted that the symplectic eigenvectors from Eq \ref{eq:10} cannot be solved using the standard eigenvalue algorithms directly. One approach uses the Schur form of skew-symmetric matrices \cite{Son2021Computing} and the details are given in the \emph{Supplementary Material} together with the corresponding Matlab script. 
\section{Discussion}
\label{sec:3}
The procedure given in the previous section tells us that there exist symplectic matrices that can decompose the Fisher information matrix (FIM). Taking analogy with the standard eigenvalue problem, in this section, we first show that the symplectic eigenvector matrix maximises a matrix trace subject to a symplectic constraint, and then demonstrate that the symplectic eigenvalues of the FIM indicate the sensitivity of the Kullback-Leibler (K-L) divergence in a symplectic basis. 
\subsection{Constrained maximization}
\label{sec:3.1}
The standard eigenvalue equation can be obtained from a trace maximization problem subject to an orthogonal constraint:
\begin{equation} \label{eq:13}
    \text{max} \: \mathrm{tr}\left(\bf{X^{\mathsf{T}}AX} \right) \quad
    \text{s.t.} \: \bf X^{\mathsf{T}}X=I
\end{equation}
To solve the constrained optimization in Eq \ref{eq:13}, the method of Lagrange Multiplier can be used. In this case, the Lagrangian and its first derivative are:
\begin{equation} \label{eq:14}
    L(\bf X,\Lambda)=\mathrm{tr} (\bf X^{\mathsf{T}}AX) - 
        \mathrm{tr} \left[ \bf \Lambda(X^{\mathsf{T}}X)-I) \right]
\end{equation}
\begin{equation} \label{eq:15}
    \frac{\partial L(\bf X,\Lambda)}{\partial \mathbf{X}} 
    = \bf AX + A^{\T}X - X(\Lambda + \Lambda^{\T}) 
    = 2\mathbf{AX}-2\mathbf{X\Lambda}
\end{equation}

where the matrix $\mathbf{\Lambda}$ is the Lagrange Multiplier and we have assumed the matrix $\mathbf{A}$ is symmetric. In addition, since the constraint $\bf X^{\T} X$ is symmetric, the Lagrange multiplier $\mathbf{\Lambda}$ is also symmetric \cite{Wen2013feasible}. Setting the first order optimality condition for the Lagrangian, the orthogonality constraint leads to the standard eigenvalue problem $\bf AX=X\Lambda$, as Eq (6) when the Fisher information matrix $\mathbf{F}$ is the symmetric matrix of interest. \\
Taking the analogy, the maximization problem can be formulated subject to a symplectic constraint \cite{Son2021Computing} :
\begin{equation} \label{eq:16}
    \text{max} \: \mathrm{tr}\left(\bf{X^{\mathsf{T}}AX} \right) \quad
    \text{s.t.} \: \bf X^{\mathsf{T}}X=J
\end{equation}
\begin{equation} \label{eq:17}
    \frac{\partial L(\bf X,\Lambda)}{\partial \mathbf{X}} 
    = \bf AX + A^{\T}X - JX\Lambda - J^{\T}X\Lambda^{\T}) 
    = 2 \mathbf{AX}-2 \mathbf{JX\Lambda}
\end{equation}

where both matrix $\mathbf{J}$ and $\mathbf{\Lambda}$ are skew symmetric, i.e. $\bf J^{\T} =-J$. The optimality condition then results the following symplectic eigenvalue problem: 
\begin{equation} \label{eq:18}
    \bf AX = JX\Lambda
\end{equation}
which has exactly the same form as Eq \ref{eq:10}. 

The standard and the symplectic eigenvectors thus provide the directions to maximise the matrix trace in an orthogonal and a symplectic basis respectively, and the corresponding eigenvalues indicate the sensitivities. A special case with the Fisher information matrix (FIM) links its eigenvalues to the sensitivities of the Kullback-Leibler (K-L) divergence, aka relative entropy, and this is described in the next section below. 
\subsection{Sensitivity of entropy}
\label{sec:3.2}
As mentioned in the introduction, consider a general function $\bf y=h(x)$, the probabilistic sensitivity analysis characterise the uncertainties of the output $\bf y$ that is induced by the random input $\bf x$. When the joint probability distribution of the output is known, the entropy of the uncertainty can be estimated as \cite{Cover2006Elements}: 
\begin{equation} \label{eq:19}
    H = - \int \pyb \ln \pyb d\mathbf{y}
\end{equation}

The perturbation of the entropy, defined as a relative entropy quantified using the K-L divergence, can be approximated by a quadratic form in the Fisher information matrix \cite{Yang2022Digital}:
\begin{equation} \label{eq:20}
    \Delta H \equiv KL \left[ \pyb || \pybb \right] =
        \int \pyb \ln \frac{\pyb}{\pybb} d\mathbf{y} \approx
        \frac{1}{2}\Delta \mathbf{b}^{\T}\mathbf{F} \Delta \mathbf{b}
\end{equation}

where the perturbed probability is approximated using its second order Taylor expansion (e.g., see the appendix of \cite{Yang2022information}). It is noted in passing that, even without the quadratic approximation to entropy, the Fisher information can be used to quantify the distribution perturbation in its own right \cite{gauchy_information_2022}. \\
Consider the standard eigenvalue decomposition of the FIM and substitute Eq \ref{eq:5} into the expression for the relative entropy in Eq \ref{eq:20}:
\begin{equation} \label{eq:21}
    2\Delta H = \Delta \mathbf{b}^{\T}\mathbf{F} \Delta \mathbf{b}
              = (\mathbf Q^{-1}\Delta \mathbf{b})^{\T}\mathbf{\Lambda} (\mathbf Q^{-1}\Delta \mathbf{b})
              = \sum_j^{2n}\lambda_j\xi_j^2
              = \sum_j^{n}\lambda_j\xi_j^2 + \lambda_{n+j}\xi_{n+j}^2
\end{equation}
where $\xi_j=(\mathbf Q^{-1}\Delta \mathbf{b})_j$ and it is clear that the eigenvalues $\lambda_j$ indicate the magnitude of the entropy sensitivity. It can be seen from Eq \ref{eq:21} that the relative entropy in this quadratic form can be regarded as an ellipsoid geometrically, i.e., $\sum\lambda_j\xi_j^2=1$ . This is a consequence of the semi-positive definiteness of the FIM and the ellipsoid is proper when the FIM is positive-definite. The eigenvectors of the FIM define the principal axes and the inverse of the square roots of the corresponding eigenvalues, i.e., $1/{\sqrt{\lambda_j}}$, are the principal radii of the ellipse. Since the principal axes are orthogonal to each other, there is no direct relationship between any pair of coordinates, say $(\xi_j, \: \xi_{n+j})$, even they are dominated by the two-parameter pairs for the same variable of interest as discussed in the introduction.  \\
Similarly, the relative entropy in the symplectic basis can be expressed as:
\begin{equation} \label{eq:22}
    2\Delta H = \Delta \mathbf{b}^{\T}\mathbf{F} \Delta \mathbf{b}
              = (\mathbf S^{-1}\Delta \mathbf{b})^{\T}\hat{\mathbf{D}} (\mathbf S^{-1}\Delta \mathbf{b})=
         \begin{bmatrix}
             \bm{\upalpha}^{\T} & \bm{\upbeta}^{\T}
         \end{bmatrix}
         \begin{bmatrix}
            \mathbf{D} &  \\
   		    & \mathbf{D} 
   		 \end{bmatrix}
         \begin{bmatrix}
             \bm{\upalpha} \\
             \bm{\upbeta}         
         \end{bmatrix}
         = \sum_j^n d_j(\alpha_j^2 + \beta_j^2)             
\end{equation}
where $\alpha_j=(\mathbf S^{-1}\Delta \mathbf{b}))_j$ and $\beta_j=(\mathbf S^{-1}\Delta \mathbf{b}))_{j+n}$. In contrast to Eq \ref{eq:21}, it can be seen that the coordinate pair $(\alpha_j, \: \beta_j)$ is now forced to form a circle with radius $1/{\sqrt{d_j}}$. The consequence is that if  $(\alpha_j, \: \beta_j)$ corresponds to the two-parameter pairs of interest, they are symplectically equivalent, in analogy to the conjugate pair, position and momentum, in Hamiltonian mechanics. 
\section{Benchmark study}
\label{sec:bench}
The Fisher information has been introduced in \cite{Yang2022Digital} for sensitivity analysis with respect to distribution parameters. A benchmark study for Fisher sensitivity, using a linear function with decreasing coefficients and a product function with constant coefficients, has been conducted in \cite{Yang_SAframework_2022}. In this section, we apply the Fisher sensitivity analysis, using both standard eigenvalue decomposition and the proposed symplectic decomposition, to a high dimensional function: 
\begin{equation} \label{eq:bench1}
     f(\mathbf{x}) = \mathbf{a}_1^{\T} \mathbf{x} + \mathbf{a}_2^{\T} \sin(\mathbf{x}) + \mathbf{a}_3^{\T} \cos(\mathbf{x}) + \mathbf{x}^{\T} \mathbf{M} \mathbf{x}   
\end{equation}
This function has a 15 dimensional input vector $\mathbf{x}$ and has been used in \cite{oakley_probabilistic_2004} for variance based sensitivity analysis. This function's coefficients, $\mathbf{a}_1, \mathbf{a}_2$ and $\mathbf{a}_3$, are chosen so that first five of the input variables have almost no effect on the output variance, $x_6$ to $x_{10}$ have a much larger effect, and the remaining five contribute significantly to the output variance. All input variables are assumed to be independent and from a standard Gaussian distribution, i.e., $x \sim \mathbb{N}(0, 1)$.

\begin{figure}[!h]
	\centering
	\includegraphics[width=10cm]{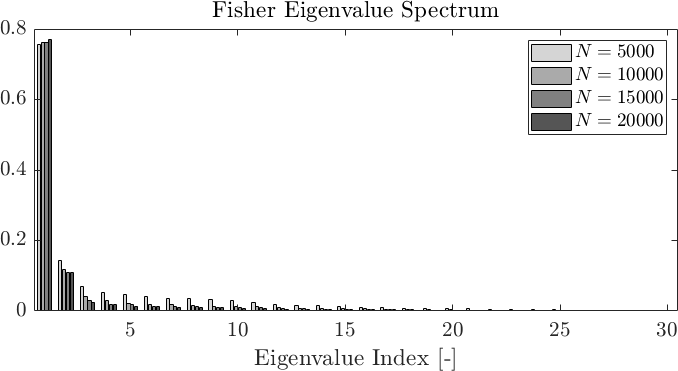}
	\caption{Eigenvalue spectrum of the FIM, for the sensitivity of the benchmark function in Eq \ref{eq:bench1}. Results are given for different number of MC samples for convergence check.}
	\label{fig:BM_Eig}
\end{figure}

\begin{figure}[!h]
	\centering
	\includegraphics[width=12cm]{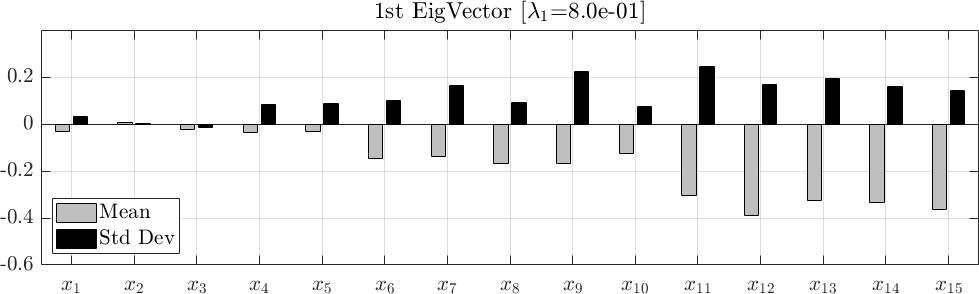}
	\caption{The first eigenvector of the FIM for the benchmark function Eq \ref{eq:bench1} with respect to distribution parameters, mean and standard deviation (Std Dev) in this Gaussian input case, of the 15 input variables. Only the dominant first eigenvector, as seen in Figure \ref{fig:BM_Eig}, is shown here.}
	\label{fig:BM_EigV}
\end{figure}

The results from the standard eigenvalue analysis of the Fisher information matrix are shown in Figure \ref{fig:BM_Eig} and Figure \ref{fig:BM_EigV}. The eigenvalue spectrum has been computed using different number of Monte Carlo samples for convergence check and the eigenvector results in Figure \ref{fig:BM_EigV} are from 20000 samples. Only the 1st eigenvector is shown in Figure \ref{fig:BM_EigV} as the eigenvalues corresponding to the rest of the eigenvectors are of much smaller amplitudes as seen in Figure \ref{fig:BM_Eig}. The sensitivity to the mean parameters of the input variables in Figure \ref{fig:BM_EigV} clearly indicates that there are there importance groups, $x_{11}$ to $x_{15}$ being the most important and $x_{1}$ to $x_{5}$ being the least important. This is in good agreement to \cite{oakley_probabilistic_2004} from a variance-based sensitivity analysis. The sensitivity to the standard deviations, on the other hand, does not show a clear clustered trend, although it is clear that the first few variables have almost no effect. Different from the variance-based analysis where only the amplitudes of the importance are measured, the Fisher sensitivity vectors also provide the relative phases of the sensitivity to the distribution parameters. For example, in Figure \ref{fig:BM_EigV}, it is clear that the effects of the input mean parameters on the output PDF uncertainty are in opposite directions to the effects due to the perturbation of the standard deviations. Note that the absolute sign of the eigenvector is arbitrary. 

\begin{figure}[!h]
	\centering
	\includegraphics[width=10cm]{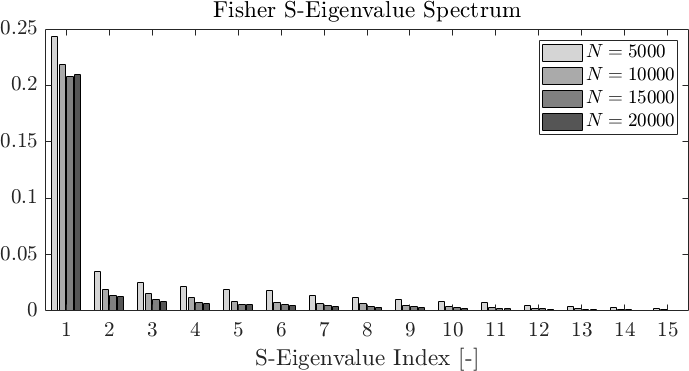}
	\caption{The symplectic eigenvalue spectrum of the FIM (S-Eig), for the sensitivity of the benchmark function in Eq \ref{eq:bench1}. Results are given for different number of MC samples for convergence check. Note that the dimension of symplectic spectrum is 15, which is half of the size of the standard eigenvalue spectrum shown in Figure \ref{fig:BM_Eig}.}
	\label{fig:BM_SEig}
\end{figure}

\begin{figure}[!h]
	\centering
	\includegraphics[width=12cm]{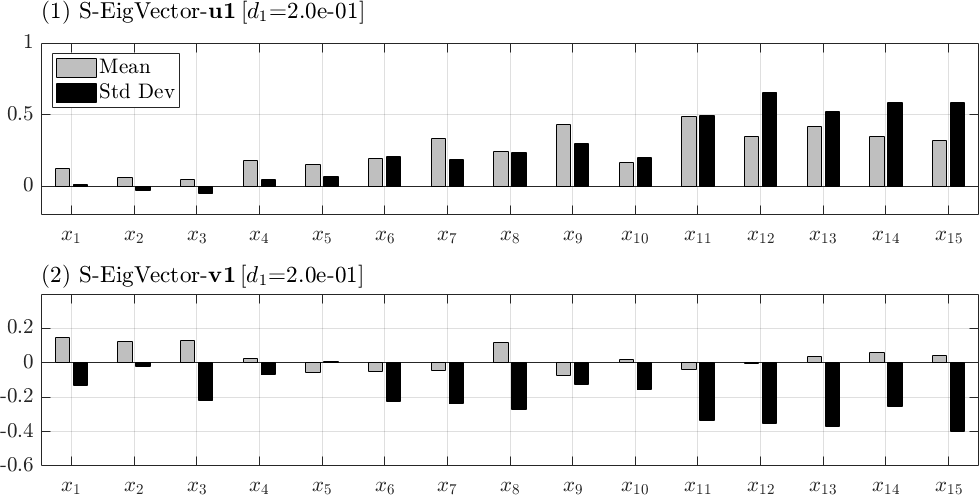}
	\caption{The first pair of symplectic eigenvectors (S-EigVector) of the FIM for the benchmark function Eq \ref{eq:bench1} with respect to distribution parameters, mean and standard deviation (Std Dev) in this Gaussian input case, of the 15 input variables. Only the dominant first pair, as seen in Figure \ref{fig:BM_SEig}, is shown here}
	\label{fig:BM_SEigV}
\end{figure}

The symplectic analysis of the FIM, for the benchmark function in Eq \ref{eq:bench1}, is shown in Figure \ref{fig:BM_SEig} and Figure \ref{fig:BM_SEigV}, in similar format to the results presented above for the standard analysis. Different from the standard eigenvalue results, the symplectic eigenvalue spectrum has a dimension half of the standard one but the symplectic eigenvectors always come in pairs. The symplectic sensitivity results in Figure \ref{fig:BM_SEigV} present a similar picture as the standard results, especially that the sensitivity to the standard deviations for $\mathbf{u1}$ vector indicate clear group importance as discussed above. However, in this case, the symplectic results do not provide any new insights. This is because the FIM is dominated by its first eigenvector in this case. For this benchmark function, there is no normalisation required for the sensitivity analysis as all input variables are equivalent. As a result, there is no compression of the sensitivity information as discussed in the motivating example and the engineering example to be studied in the Section \ref{sec:4}. 

\begin{figure}[!h]
	\centering
	\includegraphics[width=12cm]{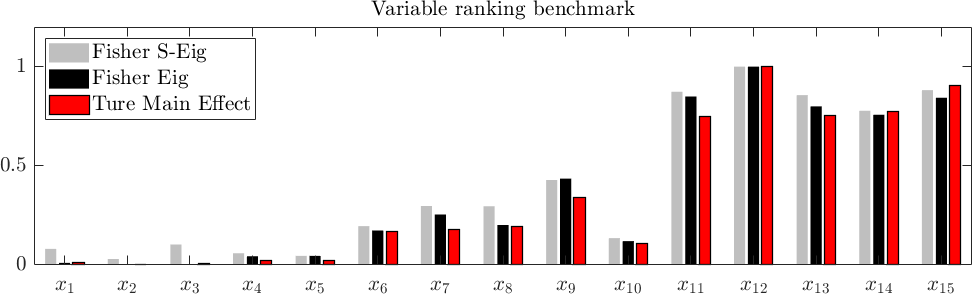}
	\caption{Variable importance ranking using three different indices: the 1st set of S-EigVector, the 1st Eigvector and the true main effect indices given in \cite{oakley_probabilistic_2004}. The results are normalised by the largest value.}
	\label{fig:BM_ParRank}
\end{figure}

For the purpose of benchmarking, the sensitivity vectors from the Fisher analysis can be further summarised for the individual variables and the results are compared to the true main effect indices given in \cite{oakley_probabilistic_2004}. The sensitivity vectors from the FIM provide the principal directions for a simultaneous variation of the input parameters. To look at the effect of individual parameters, Eq \ref{eq:21} and Eq \ref{eq:22} can be used. For example, assuming only parameter $b_k$ is varied, from Eq \ref{eq:21}:
\begin{equation} \label{eq:bench2}
         2\Delta H (\Delta b_k)  = \sum_j^{2n}\lambda_j (q_{kj} \Delta b_k)^2 = \left[\sum_j^{2n}\lambda_j q_{kj}^2 \right] \Delta b_k^2
\end{equation}
where $q_{jk}$ is the $j_{th}$ element of the eigenvector $\mathbf{q}_k$. The term inside the square bracket can be regarded as the contribution to the entropy change due to perturbation of $b_k$ alone. And the contributions from the mean and the standard deviation parameter can be further aggregated for the corresponding variables, assuming the the perturbations are independent. The resulted relative importance of the variables from the Fisher analysis, using the dominant first eigenvector (Fisher Eig) and the first pair of symplectic eigenvector (Fisher S-Eig), can then be compared to the variance based main effects and the comparison is shown in Figure \ref{fig:BM_ParRank}. Although there are small deviations, the relative importance of the three group of variables and the order of difference are clearly identified from the Fisher sensitivity analysis. Furthermore, the ratio of the first eigenvalue to the sum of all eigenvalues, as seen in Figure \ref{fig:BM_Eig}, is about 0.75 in this case and that can be regarded as the contribution of the first eigenvector to the entropy change. Although not directly comparable, 0.75 is similar to the 72$\%$ main effects contribution to the output variance as reported in \cite{oakley_probabilistic_2004}. This offers plausible suggestion that, in this case, the dominant first eigenvector of the FIM corresponds to the main effects from variance based sensitivity analysis.  

It should be noted although the contributions from perturbation of individual parameters are useful for benchmarking against variance based main effects, the purpose of the Fisher sensitivity analysis is to look at the simultaneous variations of the input parameters. Contrast to the total effects from variance based analysis, the eigenvectors and symplectic eigenvectors of FIM provide principal sensitivity directions based on the impact on the output PDF, or more specifically the entropy of the output uncertainty. Not only do the eigenvectors indicate the relative amplitude, they also provide the relative phase information of the input parameter variations as discussed earlier.  

In the next section, we will consider an engineering example where the input variables are normally of different units. As their value tend to be of different order of magnitude, normalisation is required for the Fisher sensitivity analysis. In addition, different from the scalar output from this benchmark function, engineering problems tend to have multiple outputs as the example given below. 

\section{Demonstrating application to an engineering example}
\label{sec:4}
\begin{figure}[!h]
	\centering
	\includegraphics[width=8cm]{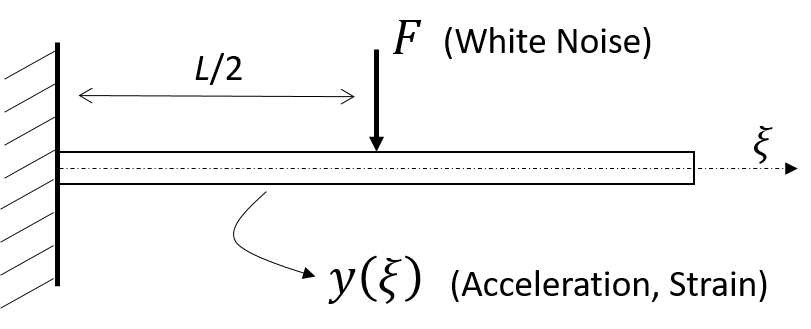}
	\caption{A cantilever beam subject to white noise excitation of unit amplitude; the responses consists of peak r.m.s acceleration and strain response. ‘peak’ indicates the maximum response along the beam for each sample of the random input. The two type of responses are normalised by the maximum values across the ensemble of the random samples.}
	\label{fig:1}
\end{figure}
In this section, we consider an engineering design example where the Fisher information is used for parametric sensitivity analysis of a cantilever beam. The beam is subject to a white noise excitation at the middle span position, see Figure \ref{fig:1}, where the excitation is bandlimited and only the first three modes are excited. In this case, the quantities of interest are the peak r.m.s responses, i.e., the maximum response along the beam, for both acceleration and strain (output $\bf y$  in Eq \ref{eq:19} is 2-dimensional). The frequency response functions for both acceleration and strain responses, at different positions along the beam, are obtained via modal summation and the modal damping is assumed to be 0.1 for all modes, see the \emph{Supplementary Material} for details. 

It is assumed that five input variables are random, $\mathbf{x} =(E,\rho,L,w,t)$, and can be described by Normal distributions, $\mathbf{b} = (\mu_m,\: \sigma_m ), \: m=1,2,\dots,5$, as listed in Table \ref{tab:1}. Two cases with different parameter values for the standard deviations of the input random variables are considered, case-1 with small variance and case-2 with big variance, with the mean values same for both cases.
To estimate the Fisher information matrix (FIM) in Eq \ref{eq:1}, an efficient numerical method based on Monte Carlo sampling and the Likelihood Ratio method is use here. The Likelihood Ratio (aka score function) method obtains a gradient estimation of a performance measure w.r.t continuous parameters in a single simulation run. More details of the method can be found in \cite{Yang2022Digital}. 

For the numerical results below, the FIM is normalised by the standard deviations (for completeness, the results with proportionally normalised FIM are given in the \emph{Supplementary Material}):
\begin{equation} \label{eq:23}
    \mathbf{F}_{\text{nor}} = \sigma_m \sigma_n \mathbf{F}_{jk} 
\end{equation}
where $j,k=1,2,\dots,10$ and $m=j/2$, $n=k/2$ when $j,k$ are even, and $m=(j+1)/2$, $n=(k+1)/2$ when $j, \: k$ are odd numbers. As discussed in the introduction, the normalisation is necessary for practical applications as the input variables are of different units and often differ by orders of magnitude. Moreover, it largely improves the condition number of the FIM. For example, in this case study, the condition number of the FIM in the order of $10^{27}$ for both case-1 and case-2, and it reduces to the order of $10^2$ for both cases after normalisation.

\begin{figure}%
    \centering
    \subfloat[\centering standard eigenvectors (EigVector)]{{\includegraphics[width=7cm]{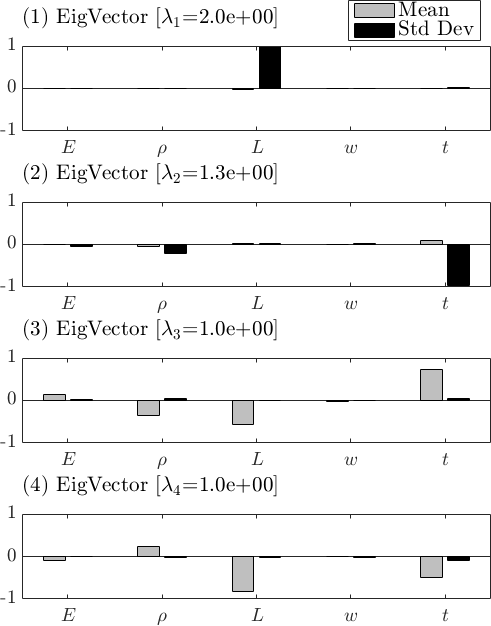} }}%
    \qquad
    \subfloat[\centering symplectic eigenvectors (S-EigVector)]{{\includegraphics[width=7cm]{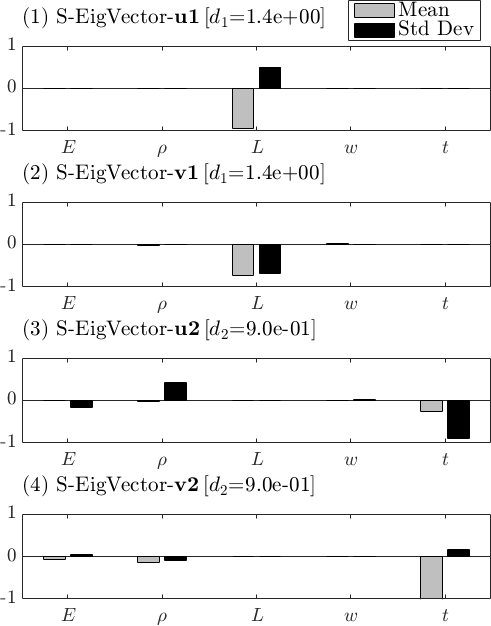} }}%
    \caption{Eigenvectors (EigVector) and symplectic eigenvectors (S-EigVector) of the FIM for Case-1. The symplectic eigenvectors come in pairs, $\bf u_1,v_1$ and $\bf u_2,v_2$ , and each pair corresponds to the same symplectic eigenvalue}%
    \label{fig:3}%
\end{figure}

\begin{figure}%
    \centering
    \subfloat[\centering standard eigenvectors (EigVector)] {{\includegraphics[width=7cm]{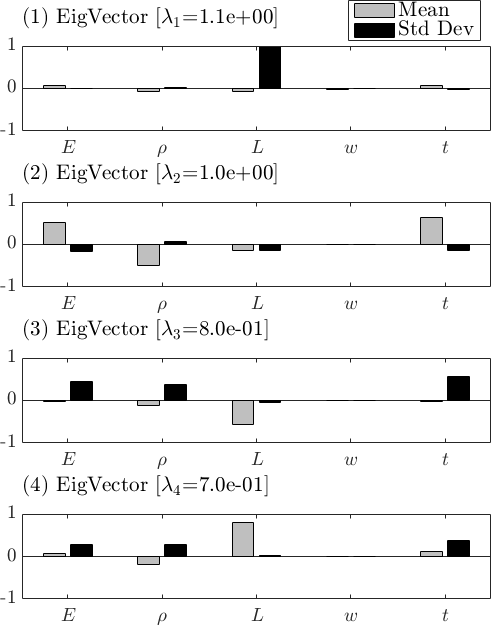} }}%
    \qquad
    \subfloat[\centering symplectic eigenvectors (S-EigVector)]{{\includegraphics[width=7cm]{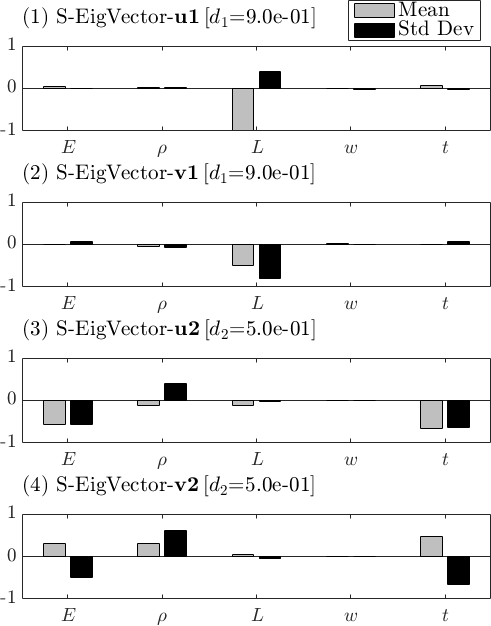} }}%
    \caption{Case-2, same key as Figure \ref{fig:3}}%
    \label{fig:4}%
\end{figure}

\begin{table}
	\caption{Mean ($\mu$) and Coefficient of Variation (CoV) for the random variables. CoV different for the two cases considered. 20000 Monte Carlo samples are used for both Case-1 and Case-2.}
	\centering
\makebox[\textwidth][c] {
	\begin{tabular}{cccccc}
		\toprule
		             & Young’s Modulus	& Density &	Length	 & Width &	Thickness \\
		\cmidrule(lr){2-6}
	               & $E [Pa]$	&  $\rho [kg/m^3]$	  & $L [m]$	  & $w [m]$	 & $t [m]$ \\
		\cmidrule(lr){2-6}
		     $\mu$ (mean) &  	69e9	 & 2700	& 0.45 &	2e-2	&  2e-3       \\
	  \midrule	    
	              &	\multicolumn{5}{c}{Case-1}                   \\
	  \cmidrule(lr){2-6}            
			$\sigma⁄\mu$ (CoV)	& 1/200	& 1/80 &	1/100	& 1/60	 &1/80    \\
		\midrule	  	
		           & \multicolumn{5}{c}{Case-2}                   \\
	   \cmidrule(lr){2-6}             	
		  $\sigma⁄\mu$ (CoV)	&1/5 &	1/5 &	1/30 &	1/6 &	1/8 \\
		\bottomrule
	\end{tabular}
	}
	\label{tab:1}
\end{table}
\begin{figure}[!h]%
    \centering 
    {{\includegraphics[width = 10cm]{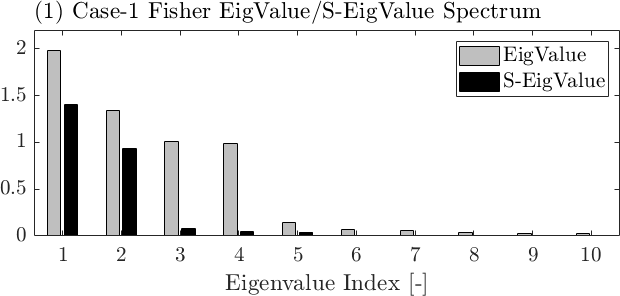} }}%
   \\[2\baselineskip]
    \centering 
    {{\includegraphics[width = 10cm]{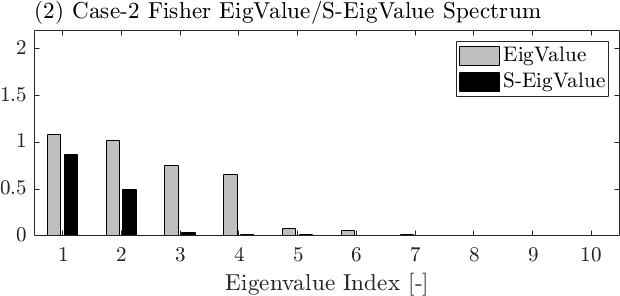} }}%
    \caption{Eigenvalues (Eig) and Symplectic eigenvalues (S-Eig) of the FIM for (1) Case-1 ; (2) Case-2}%
    \label{fig:2}%
\end{figure}
Once the FIM is estimated and normalised, the standard approach is to compute the eigenvalues and eigenvectors of the FIM for sensitivity analysis \cite{Yang2022Digital}. As discussed, the eigenvalues of the FIM represent the magnitudes of the sensitivities with respect to simultaneous parameter variations, and the most sensitive directions are given by the eigenvectors corresponding to largest eigenvalues. The standard eigenvalues and eigenvectors are denoted as ‘EigValue’ and ‘EigVector’ and are shown in Figure \ref{fig:2}, Figure \ref{fig:3} (a), Figure \ref{fig:4} (a). 

In Figure \ref{fig:2}, the eigenvalues are ordered from large to small, with the first four of similar magnitude but much larger than the rest. Note that the spectrum here is quite different from the benchmark case shown in Figure \ref{fig:BM_Eig} where one eigenvalue dominates. The corresponding first four eigenvectors are displayed in Figure \ref{fig:3} and Figure \ref{fig:4}. The results for case-1 in Figure \ref{fig:3} (a) are relatively straightforward as it can be seen that the influential parameters are $\sigma_L$, $\sigma_t$, $ \mu_L$, $\mu_t$ from the first to the fourth eigenvectors. As the first eigenvalue is almost two times of the second one, this implies we can focus on the parameter $\sigma_L$ to change the entropy the most, as discussed in Section \ref{sec:3.2}. However, the sensitivity information from the eigenvectors for case-2 in Figure \ref{fig:4} (a) are less clear. This is because the four eigenvalues in this case are of very similar magnitude. Furthermore, there are three different parameters that are important for the 2nd eigenvector, $\mu_E$, $\mu_\rho$, $\mu_t$, and the 4th eigenvector, $\sigma_E$, $\sigma_\rho$, $\sigma_t$. 

If we take a closer look at the eigenvector results in Figure \ref{fig:3} (a) and Figure \ref{fig:4} (a), it seems that there is a split phenomenon between the mean and standard deviation of the same variables. In Figure \ref{fig:3} (a), the 1st and 2nd eigenvectors point us to the standard deviations of the variables $L$ and $t$, while it is the mean values of the two variables that are important for the 3rd and 4th eigenvectors. Similarly, in Figure \ref{fig:4} (a) for case-2, $\sigma_L$ and $\mu_L$, the mean and standard deviation of the variable $L$, dominates the 1st and the 3rd eigenvector respectively, while the mean and standard deviation of $E,\: \rho, \: t$ are the influential parameters for the 2nd and 4th eigenvector. In other words, the dominance of the sensitivity to the mean and the standard deviation of the same variable splits into different eigenvectors, e.g.,  $\sigma_L$ dominates the 1st eigenvector while $\mu_L$ dominates the 3rd. 

This split phenomenon can be understood as a consequence of the normalisation as mentioned for the motivating example in Section \ref{sec:1.2}. The normalisation compresses the relative magnitudes between the eigenvalues so that the Fisher information matrix is better conditioned. This makes the ellipsoid for the relative entropy (c.f. Section \ref{sec:3.2}) closer to a sphere. The orthogonality of the principal axes could then result a split between the mean and standard deviation parameters of the same variable as their influences are similar. As a result, it is difficult to identify the most influential variables. On the contrary, the symplectic decomposition enforces a symplectic structure that tends to mitigate this issue by making the sensitivity information more distinctive by pairing the parameters of the same variable. 

As described in Section \ref{sec:2}, the same Fisher information matrix (FIM) can also be decomposed onto a symplectic basis. The symplectic eigenvalues and eigenvectors are named as ‘S-EigValue’ and ‘S-EigVector’ and are shown in Figure \ref{fig:2}, (a)-2 and (b)-2, Figure \ref{fig:3} (b), Figure \ref{fig:4} (b). The dimension of the symplectic eigenvalue spectrum, 5 in this case, is half of the standard eigenvalues. The symplectic eigenvectors come in pairs, $(\bf u_1,v_1)$ and $(\bf u_2,v_2)$ , as shown in Figure \ref{fig:3} (b), Figure \ref{fig:4} (b), and each pair corresponds to the same symplectic eigenvalue. As compared to the standard eigenvectors, the split parameters are grouped together in the symplectic eigenvectors. For example, for case-1 results in Figure \ref{fig:3} (b), the 1st symplectic eigenvector pairs identify $L$ as the influential variable, with its mean and standard deviation dominates $\bf u_1$ and $\bf v_1$ respectively. The same can be found for the variable $t$ for 2nd pair of symplectic eigenvectors $(\bf u_2,v_2)$ in Figure \ref{fig:3} (b). Similar conclusion can be found for the case-2 results in Figure \ref{fig:4} (b). The grouping of the parameters is a consequence of the symplectic structure, where parameters are regarded as two-parameter pairs, e.g., $(\mu,\sigma)$ in this case. This is really pertinent in the sensitivity analysis as it makes the influential variables, or two-parameter pairs, very distinctive.

\begin{figure}%
    \centering
    \subfloat[\centering S-EigVector ($L$ and $t$ pairing)]{{\includegraphics[width=7cm]{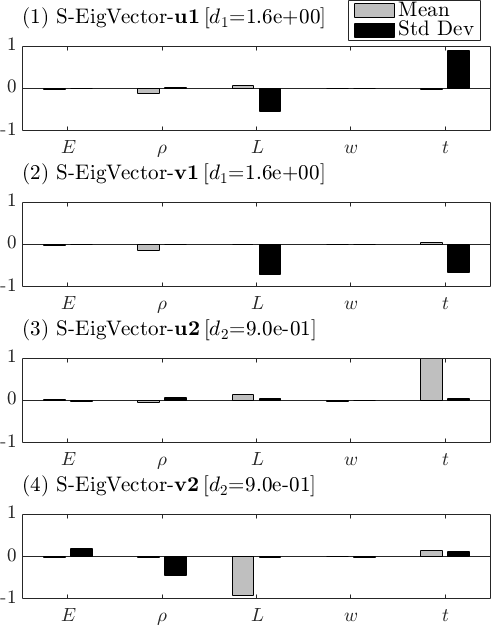} }}%
    \qquad
    \subfloat[\centering S-EigVector ($L$ and $w$ pairing)]{{\includegraphics[width=7cm]{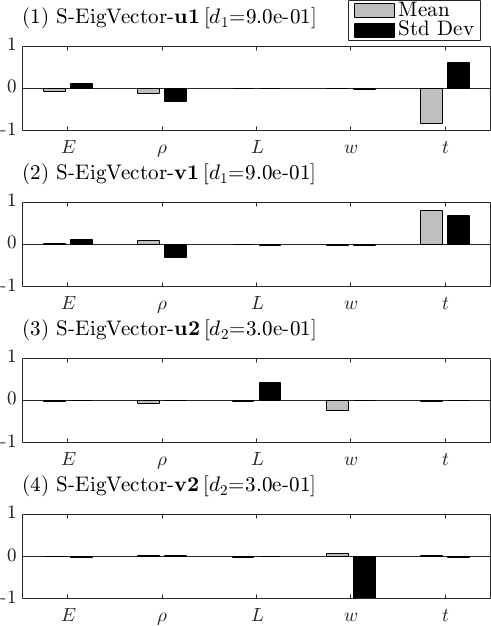} }}%
    \caption{Symplectic eigenvectors (S-EigVector) of the FIM for Case-1, for two different pairing decisions. (a) ($\mu_L$,$\mu_t$) and ($\sigma_L$,$\sigma_t$) in pairs; (b) ($\mu_L$,$\mu_w$) and ($\sigma_L$,$\sigma_w$) in pairs. The rest of the pairs are same as Figure \ref{fig:3} }%
    \label{fig:5}%
\end{figure}

It is interesting to note that in both cases, the square of the 1st symplectic eigenvalue is almost the same as the product of the two standard eigenvalues that split. For example, for case-1, $d_1^2=1.96$ and that is about the same as the product $\lambda_1 \times \lambda_3 = 1.97$, which corresponds to the 1st and 3rd eigenvectors that are dominated by $\sigma_L$ and $\mu_L$ respectively. This is a consequence of Eq \ref{eq:10c}, where when two of the standard eigenvectors splits, the product of their eigenvalues tends to be conserved in the corresponding symplectic decomposition. This also occurs for decision oriented pairings to be presented in Figure \ref{fig:5}. For case-1, the 1st and 2nd eigenvectors are dominated by $\sigma_L$ and $\sigma_t$. When these two parameters are paired together, as to be seen in Figure \ref{fig:5} (a), $d_1^2=2.56$ is very similar to the product $\lambda_1 \times \lambda_2 = 2.6$ in Figure \ref{fig:3} (a). 

Although in this simple example the parameter split found from the standard eigenvalue analysis can be easily identified, it will get more difficult with a larger number of parameters. On the contrary, the parameter pairing structure is enforced by the symplectic decomposition. The same conclusion can also be made for the proportionally normalised FIM as presented in the \emph{Supplementary Material}.  

In addition, contrary to the standard eigenvalue decomposition where the sensitivity information is fixed for a given FIM, the symplectic variant takes account of user input for the pairing decisions. As an example, two different pairing decisions are considered for the same FIM from case 1 presented in Figure \ref{fig:3}. The symplectic eigenvectors are shown in Figure \ref{fig:5}, with the rows and columns of the FIM rearranged as per the pairing requirement. It should be noted that while the standard eigenvalue analysis is invariant with respect to the row/column operation, the symplectic spectra are different as shown in Figure \ref{fig:5}. 

Instead of using the mean and stand deviation as natural pairs for the same variables, we consider pairing the mean and standard deviation parameters for two different variables. In Figure \ref{fig:5} (a), we pair $L$ and $t$, i.e., ($\mu_L$,$\mu_t$) and ($\sigma_L$,$\sigma_t$) in pairs, while in Figure \ref{fig:5} (b), we pair $L$ and $w$, i.e., ($\mu_L$,$\mu_w$) and ($\sigma_L$,$\sigma_w$) in pairs. It is noted in passing that although mainly for demonstrating purposes, these pairing decisions can arise in practice where the actions to reduce the uncertainties of two independent variables   can impact both. For example, modifying of the production line can have the same effect on the uncertainties of the length $L$ and the thickness $t$, and this would prompt a decision-oriented sensitivity analysis with respect to the parameter pairs. It is clear from Figure \ref{fig:5} that the sensitivity to the paired parameters are grouped together as before. For example, the sensitivity to the pair ($\sigma_L$, $\sigma_t$) dominates the first group of the symplectic eigenvectors in Figure \ref{fig:5} (a), and ($\sigma_L$, $\sigma_w$) are grouped together in the 2nd symplectic eigenvector pair in Figure \ref{fig:5} (b). It is interesting to note that the 1st symplectic eigenvector pair in Figure \ref{fig:5} (b) is very similar to the 2nd symplectic eigenvector pair in Figure \ref{fig:3} (b). This is mostly because the pairing for $E$, $\rho$ and $t$ are the same for the two figures. However, for $L$ and $w$ pairing in Figure \ref{fig:5} (b), the dominance of $L$ seen in Figure \ref{fig:3} disappears and $t$ becomes the most sensitive variable for the case described in Figure \ref{fig:5} (b). This demonstrates that the symplectic decomposition is decision oriented, as even for the same FIM, it extracts different sensitivity information according to different pairing strategies. 

While only an engineering design example is considered here, the benefits of the symplectic decomposition are expected for general decision problems, whenever the spectral analysis of the FIM is used for sensitivity analysis. As the additional computation cost is negligible once the Fisher information matrix is obtained, the symplectic decomposition can be used in tandem with the standard eigenvalue decomposition to extract more useful sensitivity information.  
\section{Conclusions}
\label{sec:5}
A new probabilistic sensitivity metric has been proposed based on the symplectic spectral analysis of the Fisher information matrix (FIM). Contrasting to the standard eigenvalue decomposition, the symplectic decomposition of the FIM naturally identifies the sensitivity information with respect to two-parameter pairs, e.g., mean and standard deviation of a random input. The resulted symplectic eigenvalues of large magnitude, and the corresponding symplectic eigenvectors of the FIM, then reveal the most sensitive two-parameter pairs. 

Through an engineering design example using a simple cantilever beam, it is observed that the normalisation of the FIM tends to compress the relative magnitudes between the eigenvalues. Geometrically the relative entropy ellipsoid becomes near-spherical (c.f. Section \ref{sec:3.2}) due to the normalisation, and this results a split phenomenon of different distribution parameters of the same variable. It is demonstrated that the proposed symplectic decomposition can reveal the concealed sensitivity information between the parameter pairs. Contrary to the standard eigen decomposition where the sensitivity information is fixed for a given FIM, the symplectic variant takes account of user input for the pairing decisions.  As the additional computation cost is negligible once the Fisher information matrix is obtained, the symplectic decomposition can thus be used in tandem with the standard eigenvalue decomposition to gain more insight into the sensitivity information, and orient towards a better decision support under uncertainties.  

The proposed symplectic decomposition is only applicable for parameter space of even dimensions. For distribution parameters that belong to two-parameter family of probability distributions, such as the widely used location-scale families, there is a natural pairing of the parameters. For more general cases, a decision-oriented two-parameter re-parametrization of the Fisher information matrix is necessary and that is one of the future research to be explored. 
\section*{Acknowledgment}
This work has been funded by the Engineering and Physical Sciences Research Council through the award of a Programme Grant “Digital Twins for Improved Dynamic Design”, Grant No. EP/R006768. For the purpose of open access, the author has applied a Creative Commons Attribution (CC BY) licence to any Author Accepted Manuscript version arising. The author is grateful to Professor Robin Langley, University of Cambridge, for comments on an early draft of this paper and the support to publish this work.
\section*{Data availability statement}
The datasets generated during and/or analysed during the current study are available in the GitHub repository: \url{https://github.com/longitude-jyang/SymplecticFisherSensitivity}
\bibliographystyle{unsrt}
\bibliography{references}  

\end{document}